\documentclass[english]{article}
\usepackage[T1]{fontenc}
\usepackage[latin9]{inputenc}
\usepackage{textcomp}
\usepackage{amssymb}
\usepackage{graphicx}
\usepackage{esint}

\makeatletter
\newcommand{\lyxaddress}[1]{
\par {\raggedright #1
\vspace{1.4em}
\noindent\par}
}

\makeatother

\usepackage{babel}
\begin{document}

\title{A holographic charged preon model}

\author{T. R. Mongan}

\maketitle

\lyxaddress{84 Marin Avenue, Sausalito, CA 94965 USA}

\lyxaddress{tmongan@gmail.com}
\begin{abstract}
The Standard Model (SM) is a successful approach to particle physics
calculations. However, there are indications that the SM is only a
good approximation to an underlying non-local reality involving fundamental
entities (preons) that are not point particles. Furthermore, our universe
seems to be dominated by a vacuum energy/cosmological constant. The
holographic principle then indicates only a finite number of bits
of information will ever be available to describe the observable universe,
and that requires a holographic preon model linking the (0,1) holographic
bits to SM particles. All SM particles have charges 0, 1/3, 2/3 or
1 in units of the electron charge $\pm e$, so the bits in a holographic
preon model must be identified with fractional electric charge. Such
holographic charged preon models require baryon asymmetry and also
suggest a mechanism for stationary action. This paper outlines a holographic
charged preon model where preons are strands with finite energy density
specified by bits of information identifying the charge on each end.
In the model, SM particles consist of three strands with spin states
corresponding to wrapped states of the strands. SM particles in this
wrapped preon model can be approximated by preon bound states in non-local
dynamics based on three-preon Bethe-Salpeter equations with instantaneous
three-preon interactions. The model can be falsified by data from
the Large Hadron Collider because it generates baryon asymmetry without
axions, and does not allow more than three generations of SM fermions. 
\end{abstract}

\subsection*{Introduction}

The Standard Model (SM) is a successful approach to particle physics
calculations. However, there are indications that the SM is only a
good approximation to an underlying finite-dimensional non-local reality
involving fundamental entities that are not point particles. First,
the SM is a local field theory, and (consistent with quantum mechanics)
there are non-local effects in our universe {[}1{]}. Second, our universe
is apparently dominated by a vacuum energy/cosmological constant.
According to the holographic principle, this indicates only a finite
number of bits of information will ever be available to describe the
observable universe {[}2{]}. Third, energy densities of the basic
SM point particles are infinite, and infinities in physical theories
indicate inadequacies of those theories.

The holographic principle \emph{requires} a preon model to describe
SM particles in terms of the (0,1) holographic bits of information.
All Standard Model particles have charges 0, 1/3, 2/3 or 1 in units
of the electron charge $\pm e$, so the bits in a preon model must
be identified in some way with fractional electric charge. Also, in
any physical system, energy must be transferred to change the information
in a bit from one state to another. Labelling the low energy state
of a bit $\mbox{e/3n}$ and the high energy state by $-e/3n$ then
amounts to defining electric charge. If the universe is charge neutral,
as it must be if it began by a spontaneous quantum fluctuation from
nothing, there must be equal numbers of $e/3n$ and $-e/3n$ charges.
Any holographic preon model in such a universe then embodies charge
conservation, a precondition for gauge invariance and Maxwell\textquoteright{}s
equations. Furthermore, because the energy of $e/3n$ and $-e/3n$
bits are unequal, any such preon model \emph{requires} baryon asymmetry.
In a closed universe, the baryon asymmetry is readily calculated and
agrees with that necessary to produce today\textquoteright{}s matter
dominance {[}3{]}.

Since the holographic principle requires preons, and results in baryon
asymmetry, the purpose of this paper is to show that at least one
preon model can link (0,1) bits of information in a holographic model
to SM particles. The model uses two ideas from Bilson-Thompson\textquoteright{}s
preon model {[}4{]}, the concept that preons are strands instead of
point particles, and the hint that spin results from wrapped preon
strands.

\subsection*{Background on holography}

The holographic principle says all information that will ever be available
to any observer about physics \emph{within }a horizon is given by
the finite amount of information \emph{on }the horizon, specified
by one quarter of the horizon area in Planck units {[}2{]}. Consistent
with the model in {[}5{]}, the universe is dominated by vacuum energy
density $\varepsilon_{v}$ related to a cosmological constant $\Lambda\approx10^{-56}$cm$^{-2}$
by $\varepsilon_{v}=\frac{\Lambda c^{4}}{8\pi G}$, where the gravitational
constant $G=6.67\times10^{-8}$cm$^{3}$/g sec$^{2}$ and $c=3\times10^{10}$cm/sec.
Therefore, the universe is asymptotic to a de Sitter space, with an
event horizon at distance $R_{H}=\sqrt{\frac{3}{\Lambda}}=1.7\times10^{28}$cm.
The holographic principle then says only $N=\frac{3\pi}{\Lambda\delta^{2}\ln2}=5\times10^{122}$
bits of information ($\frac{3\pi}{\Lambda\delta^{2}}$ degrees of
freedom) will ever be available to describe the universe, where the
Planck length $\delta=\sqrt{\frac{\hbar G}{c^{3}}}$, and $\hbar=1.05\times10^{-27}$g
cm$^{2}$/sec. So, theories involving continuum mathematics (such
as the SM) can only approximate an underlying finite-dimensional holographic
theory.

\subsection*{Holography and non-locality}

Quantum mechanics (and the reality it describes) is known to be non-local
{[}1{]}, but the mechanism of non-locality has remained obscure. However,
a holographic quantum mechanical theory is essentially non-local if
the wavefunction on the horizon (describing the evolution of all information
available about the observable universe) is the boundary condition
on the wavefunction describing the scalar information distribution
within the horizon. Then any quantum transition altering the wavefunction
on the horizon is reflected in an instantaneous non-local change in
the wavefunction describing the distribution of information within
the horizon that describes all physics in the observable universe.

A holographic quantum mechanical description of the bits of information
on the horizon requires wavefunctions specifying the probability distribution
of those bits of information. A quantum description of the information
available to describe SM particle interactions within the horizon
can be obtained {[}6{]} by identifying each area (pixel) of size $4\delta^{2}\ln2$
on the horizon with one bit of information on the horizon at radius
$R_{H}=\sqrt{\frac{3}{\Lambda}}$. A holographic preon theory must
then relate the bits of information on the horizon to SM particles
and their constituent preons within the horizon.

\subsection*{Preons with finite energy density}

Infinities in a physical theory indicate inadequacies in the theory,
so infinite energy densities of SM particles or preons must not occur
in a reasonable holographic model. If the fundamental preon entities
within the horizon are one-dimensional strands characterized by a
bit of information on each end, both preons and SM particles made
from them have finite energy density. Furthermore, there is a straightforward
way to relate information on the horizon to information on one-dimensional
strands constituting preons within the horizon, with each strand carrying
a bit of information on each end.

Consider a wavefunction specified in a spherical coordinate system
centered on the observer's position. The z axis of such a coordinate
system pierces the $i$th pixel of area $4\delta^{2}\ln2$ on one
hemisphere of the horizon and the antipodal $j$th pixel of area $4\delta^{2}\ln2$
on the opposite hemisphere of the horizon. Suppose $\theta_{i}$ is
the polar angle measuring the angular distance from the $z$ axis
through the $i$th pixel to the point on the horizon where the wavefunction
is evaluated. Then wavefunctions of the form $\cos\theta_{i}$ on
the horizon have wavelength $2\pi R_{H}$ and can define the probability
distribution for finding the information associated with the antipodal
pair of bits at any location on the horizon, with the maximum probability
of finding information associated with those two bits in the two pixels
where the $z$ axis pierces the horizon. If the $z$ axis of a wavefunction
$\phi_{k}=\sqrt{\frac{3}{2\pi R_{H}}}\cos\theta_{i}$ is associated
with the $k$th pair of antipodal bits of information $(i,j)$, state
identifiers (0,0), (0,1), (1,0) or (1,1) specify the two bits of information
associated with the wavefunction $\phi_{k}$ for any of the $N/2$
quantum states on the horizon. The wavefunction $\phi_{k}$ for the
$k$th pair of antipodal bits of information $(i,j)$ is then identified
with the wavefunction defining the probability that the $(i,j)$ bits
of information on the ends of the $k$th strand within the horizon
lie along the axis connecting any two antipodal pixels. In an observer's
coordinate system chosen to describe all information on the horizon,
a quantum state with zero in the pixel on the horizon in one radial
direction and one in the antipodal pixel is not identical to a state
with one in the first pixel and zero in the antipodal pixel. So, quantum
states with identifiers (1,0) and (0,1) on the horizon (and the corresponding
strands) are not identical.

\subsection*{Spin and preon wrapping}

In this model, wavefunctions for quantum states on the horizon with
state identifiers (0,0), (0,1), (1,0) or (1,1) are boundary conditions
on wavefunctions specifying the probability of finding the ends of
preon strands {[}with charges $(-e/6,\:-e/6)$, $(-e/6,\; e/6)$,
$(e/6,\;-e/6)$ or $(e/6,\; e/6)$, respectively, on those ends{]}
lying along any given radial axis in the observer\textquoteright{}s
coordinate system within the horizon. When the $N/2$ wavefunctions
$\phi_{k}$ describing the $N$ bits of information in the universe
are transformed to the observer's coordinate system, their sum is
the wavefunction for the probability distribution of all information
available on the horizon. In a closed universe, given the instantaneous
probability distribution of information available on the horizon obtained
from the wavefunctions $\phi_{k}$, the instantaneous probability
distribution of preon strands within the horizon can (in principle)
be determined, as explained in Appendix A.

SM particles with spin can be represented by preon strand triads involving
information in two antipodal clusters of three adjacent pixels on
opposite hemispheres of the horizon, bound into SM particles by non-local
three-body interactions. SM particles with spin are identified with
wrapped strand triads and different strand wrapping configurations
correspond to the spin of SM particles. Strand configurations include:
a) a straight line segment along the axis between charges, b) the
open strand wrapping configurations at the top of Figure 1, with ends
on the axis between charges, and c) the strand rings at the top of
Figure 2, comprising open charged or closed neutral rings with the
ends of the strands close together along the axis between charges.
States (0,1) and (1,0) represent straight neutral strands with charges
$(-e/6,\; e/6)$ or $(e/6,\;-e/6)$ on their ends \emph{or} closed
neutral rings formed by joining oppositely charged ends of open neutral
strands. Note that a pair of $(-e/6,\;-e/6)$ and $(e/6,\; e/6)$
strands, with a net neutral charge, do not exist within a single SM
particle because they are equivalent to a pair of $(-e/6,\; e/6)$
and $(e/6,\;-e/6)$ strands. The basic SM particles are spin \textonehalf{}
fermions, with forces between them mediated by spin 1 or spin 2 bosons.
A spin zero Higgs boson can be modeled by a bound state of three straight
strands lying along mutually perpendicular axes with no preferred
direction. Charge and angular momentum conservation in the crossed
channel requires neutral or unit charged SM bosons mediating forces
between SM fermions. Two rules relate three-strand preon configurations
to SM particles with spin: 1) Spin \textonehalf{} SM fermions are
comprised of one effective right or left-handed open strand wrap around
one or two straight strands. Right-handed or left-handed wrapping
corresponds to spin up or spin down. 3) SM bosons are comprised of
one or more net closed rings wrapped around one or two straight strands
defining the spin axis of the boson.

\subsection*{Wrapped preon configurations and specific SM particles}

The $N$ bits of information on an observer's horizon constitute all
information available to describe physics within the four-dimensional
spacetime of a closed vacuum-dominated universe. In a reasonable preon
model, a set of three-strand configurations related to the $N$ bits
of information on the horizon must be identified unambiguously with
SM particles. However, it is not necessary that a physical particle
be associated with \emph{every} conceivable strand configuration.
The three possible ways one open strand wrap can occur allows three
generations of spin \textonehalf{} SM fermions. The three configurations,
shown in Figure 1 for spin up fermions, are: a) two partial open strand
wraps producing one effective full open strand wrap surrounding a
central straight strand, b) an open strand wrapping one straight strand,
accompanied by a \textquotedbl{}spectator\textquotedbl{} straight
strand, and c) an open strand wrapping two straight strands. The effective
mass of strands bound into massive SM particles with Compton wavelength
$\lambda$ is of the order of $m=\frac{h}{\lambda c}$ where $h$
is Planck\textquoteright{}s constant. Replacing various neutral strands
with charged strands results in the charge states of the three generations
of SM fermions. Denote neutral strands $(e/6,\;-e/6)$ by $A$ and
neutral strands $(-e/6,\; e/6)$ by $B$. Then neutrinos are linear
superpositions of configurations with three neutral strands $(AAA,\; AAB,\; ABB,$
and $BBB)$, charge 1/3 quarks are linear superpositions of configurations
with two neutral strands $(AA,\; AB$ or $BB)$, charge 2/3 quarks
are linear superpositions of configurations with one neutral strand
($A$ or $B$) and unit charge fermions are configurations with all
strands carrying the same charge. Note that two charged strands in
a single SM particle do not interact electromagnetically because incoming
charged strands in the cross channel of potential strand-strand electromagnetic
interactions cannot combine to make the crossed channel three strand
virtual photon intermediate state necessary to mediate electromagnetic
interactions between two charged strands.

In this model, double open strand wraps around a single straight strand
are not relevant. Two open strand wraps in the same direction around
a straight strand are topologically equivalent to one open strand
wrap in the opposite sense around two straight strands and do not
constitute new particles. Two open strands wrapped in opposite directions
around a straight strand would result in a non-physical state with
no spin but with an (unphysical) preferred axis along the straight
strand.

In Figure 2, spin one SM bosons are represented by one net closed
ring around one or two straight strands and spin two SM bosons (gravitons)
are represented by two neutral rings wrapped around one straight strand.
Therefore, SM bosons in this preon model can not have spin greater
than two, because three rings leave no remaining straight strands.
Neutral bosons involve neutral rings and non-identical straight neutral
strands, and the model allows for the spin states of all bosons that
exist as independent particles outside color neutral hadrons. A neutral
ring threaded through another neutral ring, with a straight neutral
strand ($A$ or $B$) also going through the ring, gives two photon
helicity states. Two neutral rings around a straight neutral strand
($A$ or $B$) gives two (general relativitistic) graviton helicity
states. Gluons cannot exist as independent particles outside of color
neutral hadrons, and they are discussed in the next section.

As compared to the strong, electromagnetic and gravitational forces
acting \emph{between} SM particles, weak interactions in this preon
model act \emph{within} SM particles. That is, the weak interaction
modifies the arrangement of preon strands within SM fermions to change
an up quark to a down quark, or a muon to an electron. Wavelengths
of preon strands involved in these interactions within SM particles
are generally much shorter than the wavelengths of preons involved
in interactions between SM particles. The effective mass of strands
of wavelength $\lambda$ bound in SM bosons is of order $m=\frac{h}{\lambda c}$,
and the short wavelengths involved in weak interactions within SM
fermions are associated with the high mass of $Z^{0}$ and $W$ weak
bosons. A neutral ring wrapped around two straight neutral strands
$(AA,\; AB$ or $BB)$ provides three spin states of the $Z^{0}$
boson. $W$ bosons are unique among SM bosons because they involve
three charged strands, either $[(-e/6,\;-e/6),\;(-e/6,\;-e/6)$ and
$(-e/6,\;-e/6)]$ or $[(e/6,\; e/6),\;(e/6,\; e/6)$ and $(e/6,\; e/6)]$.
SM bosons with one or two charged strands are forbidden by charge
conservation in the cross channel of SM fermion-fermion scattering.
$W$ bosons are represented by two open charged rings, forming one
effective closed ring around a straight charged strand. There are
three distinguishable $W$ boson states, corresponding to the three
spin states of the $W$ boson. The state with two intertwined open
rings is identified as the state with $s_{z}=0$. The state with an
\textquotedbl{}upper\textquotedbl{} configuration of open rings, with
the ring open towards the positive $x$ axis above the ring open toward
the negative $x$ axis, is identified as the state with $s_{z}=1$.
The state with a \textquotedbl{}lower\textquotedbl{} configuration
of open rings, with the ring open towards the positive $x$ axis below
the ring open toward the negative $x$ axis, is identified as the
state with $s_{z}=-1$. Note that the $AB$ configuration of the $Z^{0}$
boson and the spin up and spin down configurations of the $W$ boson
are not identical under reflection in a mirror, so they are not parity
invariant.

\subsection*{Color neutrality of SM particles}

Each quark state in Figure 1 has an \textquotedbl{}odd strand out.\textquotedbl{}
The odd strand out is the neutral strand in charge 2/3 quarks and
the charged strand in charge 1/3 quarks. Locating the odd strand in
one of the three different possible locations in the quark configurations
for each generation provides for color states in quantum chromodynamics.
However, specifying a color charge for a particle requires an extra
bit of information in addition to the six bits specifying the preon
strands within that particle. Because the $N$ bits of information
on the horizon comprise all information describing the universe, quarks
must only occur as parts of color neutral composites (e.g., quark-antiquark
pairs comprising mesons and color neutral three quark states comprising
baryons) and gluons must not exist independently outside of hadrons.
The gluon configuration in Figure 2 allows eight different kinds of
(virtual) gluons confined within hadrons. The eight states are states
with a neutral ring wrapped around one or the other of the straight
strands, an $A$ or $B$ strand inside the ring and an $A$ or $B$
strand as a spectator.

\subsection*{Framework for calculations involving the preon model}

In a quantum mechanical holographic theory, conditions at an observer\textquoteright{}s
position at a given instant are determined by information on the observer\textquoteright{}s
horizon at the same instant. The bits of information associated with
pixels on the horizon represent SM particles within the horizon. However,
observers can only obtain information from the horizon with a time
delay of the order of billions of years. Furthermore, information
gleaned from astronomical observations at various lookback distances
corresponds to conditions on the horizon at different epochs in the
past. So, developing a finite-dimensional lattice gas holographic
model for the interaction of pixels to describe particle interactions
within the horizon is likely to be very difficult.

We will probably always need to do particle physics calculations based
on data collected in our immediate neighborhood within the horizon.
So, the SM is likely to remain a useful tool for calculating particle
interactions for a long time. When supplementing SM calculations,
the large number of degrees of freedom in finite-dimensional holographic
representations of the universe indicates approximations involving
continuum mathematics will be required. Therefore a continuum mathematics
framework for particle physics calculations accounting for non-local
aspects of this preon model is outlined below.

Non-local interactions of the three preon strands composing SM particles
can be approximated by three-preon Bethe-Salpeter field-theoretic
equations {[}7{]} involving only non-local three-preon forces. Instantaneous
three-preon interactions (consistent with instantaneous transitions
required in a non-local holographic quantum mechanical theory), and
free particle propagators for the three strands, simplify these equations
and reduce them to six-dimensional relativistically-covariant equations
involving the six four-momenta $p_{1}$, $p_{2}$, $p_{3}$, $p'_{1}$,
$p'_{2}$, and $p'_{3}$ of incoming and outgoing states in three-strand
interactions {[}7{]}. If the universe is closed, the number of bits
of information in the universe is constant (because there is nowhere
else to serve as a source or sink of information) and the number of
bits of information in the universe is twice the (conserved) number
of strands in the universe. In that case, strand production thresholds
are not an issue when dealing with the three-preon Bethe-Salpeter
equations.

The three-preon Bethe-Salpeter equations can be partial-wave analyzed
for states of angular momentum $L$ by the method of Omnes {[}8{]}.
When the instantaneous Born term $V_{L}(p_{1},\; p_{2},\; p_{3};\; p'_{1},\; p'_{2},\; p'_{3})$
in the partial-wave Bethe-Salpeter equation has the non-local separable
form $V_{L}(p_{1},\; p_{2},\; p_{3};\; p'_{1},\; p'_{2},\; p'_{3})=g_{L}(\overrightarrow{p_{1}})\; g_{L}(\overrightarrow{p_{2}})\; g_{L}(\overrightarrow{p_{3}})\; g_{L}(\overrightarrow{p'_{1}})\; g_{L}(\overrightarrow{p'_{2}})\; g_{L}(\overrightarrow{p'_{3}})$,
the partial-wave three-preon Bethe-Salpeter equations can be solved
algebraically (like the non-relativistic Lippman-Schwinger equations
in Ref. 9), producing only one bound state configuration for each
interaction. Therefore, SM particles can be identified as bound states
of various non-local separable three strand interactions in three-preon
partial-wave Bethe-Salpeter equations. A minimum of eight non-local
separable interactions, one for each strand wrapping configuration,
is needed to produce the structures in Figures 1 and 2. The required
interactions are: three $V_{L=1/2}$ interactions (one for each three
strand wrapping configuration of the three fermion generations); four
$V_{L=1}$ interactions (one for the two charged open rings and one
straight charged strand in $W$ bosons, one for the single closed
neutral ring around two straight neutral strands of the $Z^{0}$ boson,
one for the single closed neutral ring around a closed neutral ring
and a straight neutral strand for the photon, and one for the single
closed neutral ring around a straight neutral strand accompanied by
a spectator straight neutral strand for the eight confined gluons);
and one $V_{L=2}$ interaction for the double closed neutral rings
around the straight neutral strand of the graviton. An additional
non-local separable interaction decribing a bound state of three straight
neutral strands lying along mutually perpendicular axes with no preferred
direction could accompodate a spinless Higgs boson. Nine strand non-local
vertices corresponding to three line vertices in SM Feynman diagrams
must be constrained to equal the associated SM coupling constant.
Dynamical equations involving these non-local separable interactions
will provide a continuum mathematics approximation to an underlying
finite-dimensional non-local holographic theory generating SM particles.
Corrections to the SM suggested by such an approximation might provide
evidence for an underlying non-local theory. However, specifics of
detailed calculations involving this non-local separable approximation
to the SM depend on the functional forms chosen for the form factors
$g_{L}(\overrightarrow{p})$ in the non-local Born term for three-preon
Bethe-Salpeter equation and are beyond the scope of this exploratory
paper.

\subsection*{Conclusion}

For many years, particle physics employed the abstract concept of
massive point particles, despite the mathematical complications from
infinite energy densities of point particles. Wrapped preon strands
extend the idea of a massive point particle to the concept of a strand
with a charge at each end. If the wrapped preon model discussed in
this paper reasonably represents reality, there is no room for supersymmetric
partners, fourth generation SM fermions, or free particles with color
charge. Therefore, finding them at the Large Hadron Collider (or anywhere
else) will immediately falsify the model.

\subsection*{Appendix A: Preon strand distributions in a closed universe}

Consider the relation between holographic information on the horizon
and the distribution of preon strands in a closed universe. The wavefunction
for the probability of finding a strand anywhere in the universe is
a solution to the Helmholtz wave equation in the universe, and the
wavefunction specifying the information on the horizon is the boundary
condition on that wavefunction specifying the probability distribution
of strands within the universe. A closed universe with radius of curvature
$R$ can be defined by three coordinates $\chi,\;\theta$ and $\phi$,
where the volume of the three-sphere $(S^{3})$ is $R^{3}\intop_{0}^{\pi}\sin^{2}\chi d\chi\intop_{0}^{\pi}\sin\theta d\theta\intop_{0}^{2\pi}d\phi$.
The wavefunction for two bits of information on the horizon is the
projection on the horizon at $R_{H}=R\sin\chi$ of the wavefunction
for the bits of information on the ends of a strand within the volume
of the universe. One solution to the Helmholz equation on the three-sphere
{[}10{]} is the scalar spherical harmonic $Q_{10}^{2}=\sqrt{\frac{12}{\pi}}\cos\theta_{i}\sin\chi$.
For the $(i,j)$ bits of information associated with antipodal pixels,
the $\cos\theta_{i}$ behavior of the wavefunction on the horizon
determines the wavefunction $\Psi_{k}=C_{3}Q_{10}^{2}(i,j)=C_{3}\sqrt{\frac{12}{\pi}}\cos\theta_{i}\sin\chi$
as the solution of the Helmholz wave equation describing the probability
distribution within the universe for the strand associated with the
$(i,j)$ bits. The constant $C_{3}=\sqrt{\frac{4}{\pi^{2}R^{3}}}$
is determined by normalizing the wavefunction to one strand (associated
with two of the $N$ bits of information) within the universe. When
the $N/2$ wavefunctions $\Psi_{k}$ describing the distribution of
the $N$ bits of information within the universe are transformed to
the observer's coordinate system, their sum is the wavefunction for
the probability distribution of all the information in the universe.

\subsection*{Appendix B: Local consequences of stationary action and holography}

Local field theory has been the basis of theoretical physics for many
years, and is likely to be used for a long time in the future. Therefore
it is appropiate to ask what, in principle, might be the effects of
holography on local field theory. Field equations of motion are obtained
by applying the principle of stationary action to the action expressed
as an integral over the Lagrangian density. Consider a post-inflationary
Friedmann universe that is so large it is almost flat, and fields
with a Lagrangian density $\mathcal{{\normalcolor {\normalcolor \mathcal{L}}}}$$\left(\phi_{i}(x),\partial_{\mu}\phi_{i}(x))\right)$
that is a function of the fields $\phi_{i}(x)$ and their spacetime
derivatives $\partial_{\mu}\varphi_{i}(x)$. Then the action $S$
in the neighborhood of a spacetime point is the four dimensional integral
$S$ = $\int dt\int d^{3}x$ $\mathcal{{\normalcolor {\normalcolor \mathcal{L}}}}$$\left(\phi_{i}(x),\partial_{\mu}\phi_{i}(x))\right)$.
The action integral must include the Einstein-Hilbert term to account
for gravity and the Lagrangian density for the Standard Model fields.
Field equations of motion at the space-time point are found by applying
the calculus of variations and the principle of stationary action
to the action integral.

Now consider a holographic spherical screen (HSS) at radius $R$ around
a spacetime point. The action $S$ in the neighborhood of the spacetime
point can be written as $S$ = $\int dt\int r^{2}sin\theta drd\theta d\varphi$$\mathcal{{\normalcolor {\normalcolor \mathcal{L}}}}$$\left(r,\theta,\varphi)\right)$.
The holographic principle says all information available to describe
physics within the HSS resides on the HSS. Since physics within the
HSS is governed by the principle of stationary action, the action
on the HSS must be stationary. The action on the HSS surrounding a
spacetime point is $S$ = $\int dt\int R^{2}sin\theta d\theta d\varphi$$\mathcal{{\normalcolor {\normalcolor \mathcal{L}}}}$$\left(R,\theta,\varphi)\right)$.
Holographic field equations of motion on the HSS are then found by
applying the calculus of variations and the principle of stationary
action to the action integral over the HSS. The action $S$ = $\int dt\int R^{2}sin\theta d\theta d\varphi$$\mathcal{{\normalcolor {\normalcolor \mathcal{L}}}}$$\left(R,\theta,\varphi)\right)$
of a physical system is stationary, so Padmanabhan {[}11{]} notes
``the time integral reduces to multiplication by the range of integration.''
Therefore, the action per unit time on the HSS has units of energy
and is given by $S'$ = $\int R^{2}sin\theta d\theta d\varphi$$\mathcal{{\normalcolor {\normalcolor \mathcal{L}}}}$$\left(R,\theta,\varphi)\right)$.
The average Langrangian density per unit area on the HSS is then $\mathcal{L}'=\frac{S'}{4\pi R^{2}}$.

The holographic principle {[}2{]} says all information available about
physics \emph{within} an HSS at distance $R$ from an observer is
given by the finite amount of information \emph{on} the HSS. The number
of $(0,1)$ bits of information on the HSS, specified by one quarter
of the HSS area in Planck units {[}2{]}, is $N=\pi R^{2}/(\delta^{2}\ln2)$.
The Planck length $\delta=\sqrt{\frac{\hbar G}{c^{3}}}$, where $G=6.67\times10^{-8}$
cm$^{3}$/g sec$^{2}$, $\hbar=1.05\times10^{-27}$g cm$^{2}$/sec,
and $c=3\times10^{10}$cm/sec. According to the holographic principle,
the information available in the action per unit time on the HSS must
be encoded in the $N$ bits of information available on the HSS. Therefore,
the integral $S'$ = $\int R^{2}sin\theta d\theta d\varphi$$\mathcal{{\normalcolor {\normalcolor \mathcal{L}}}}$$\left(R,\theta,\varphi\right)$
must be viewed as a continuum approximation to the finite sum $S'=\sum_{i=1}^{N}\mathcal{L}_{i}4\delta^{2}\ln2$,
where $4\delta^{2}\ln2$ is the area on the HSS corresponding to the
$i$th bit with Lagrangian density $\mathcal{L}_{i}$.

In any physical system, energy must be transferred to change information
in a bit from one state to another. If the energy required to change
the state of a bit is $2E_{d}$, $\mathcal{L}_{i}=\mathcal{L}'\pm E_{d}$
. If the energy of one bit on the HSS drops from $\mathcal{L}_{i}=\mathcal{L}'+E_{d}$
to $\mathcal{L}_{i}=\mathcal{L}'-E_{d}$ , the energy of another bit
must increase from $\mathcal{L}_{j}=\mathcal{L}'-E_{d}$ to $\mathcal{L}_{j}=\mathcal{L}'+E_{d}$.
This is the mechanism for stationary action. The relevant time intervals
are multiples of $2\sqrt{\ln2}t_{P}$, the time it takes for a light
signal to travel from one pixel of area $4\delta^{2}\ln2$ on the
HSS to the adjacent pixel. The Planck time $t_{P}=\sqrt{\frac{\hbar G}{c^{5}}}$.

The energy to change the state of a bit can't depend on the radius
of the HSS, because many HSS are possible. So, the energy to change
the state of a bit must be transferred by a massless quantum with
wavelength related to the size of the universe. It is not clear how
to relate the wavelength of the massless quantum to the scale factor
of a flat or open universe. However, the only macroscopic length characteristic
of the size of a closed Friedmann universe with radius $R_{u}$ is
the circumference $2\pi R_{u}$, so $2E_{d}=\frac{\hbar c}{R_{u}}$.

Field equations within the HSS are obtained by applying the principle
of stationary action to the volume integral over the Lagrangian density
within the HSS. Field equations on the HSS are obtained by applying
the principle of stationary action to the surface integral of the
Lagrangian density on the HSS. The solution to the field equations
on the HSS is the boundary condition on the solution to the field
equations within the HSS. Those boundary conditions, and the associated
solutions to the field equations within the HSS, change with every
time step of changes in the components of the finite action sum $S'=\sum_{i=1}^{N}\mathcal{L}_{i}4\delta^{2}\ln2$.

\begin{figure}
\includegraphics[scale=0.15]{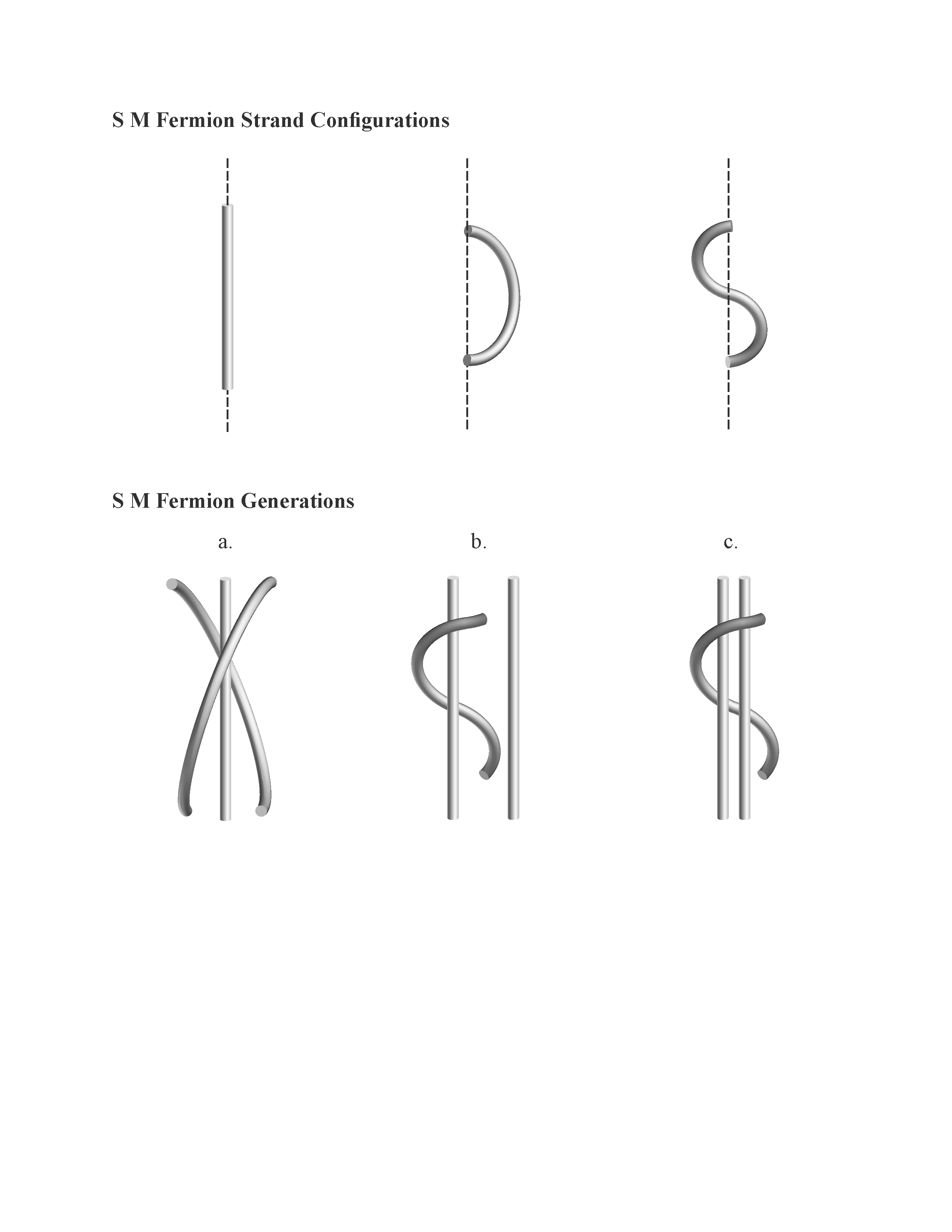}\caption{Preon strand configurations producing spin up Standard Model fermions
and the resulting three preon fermion states. Strands are shown as
tubes for ease of visualization.}
\end{figure}

\begin{figure}
\includegraphics[scale=0.15]{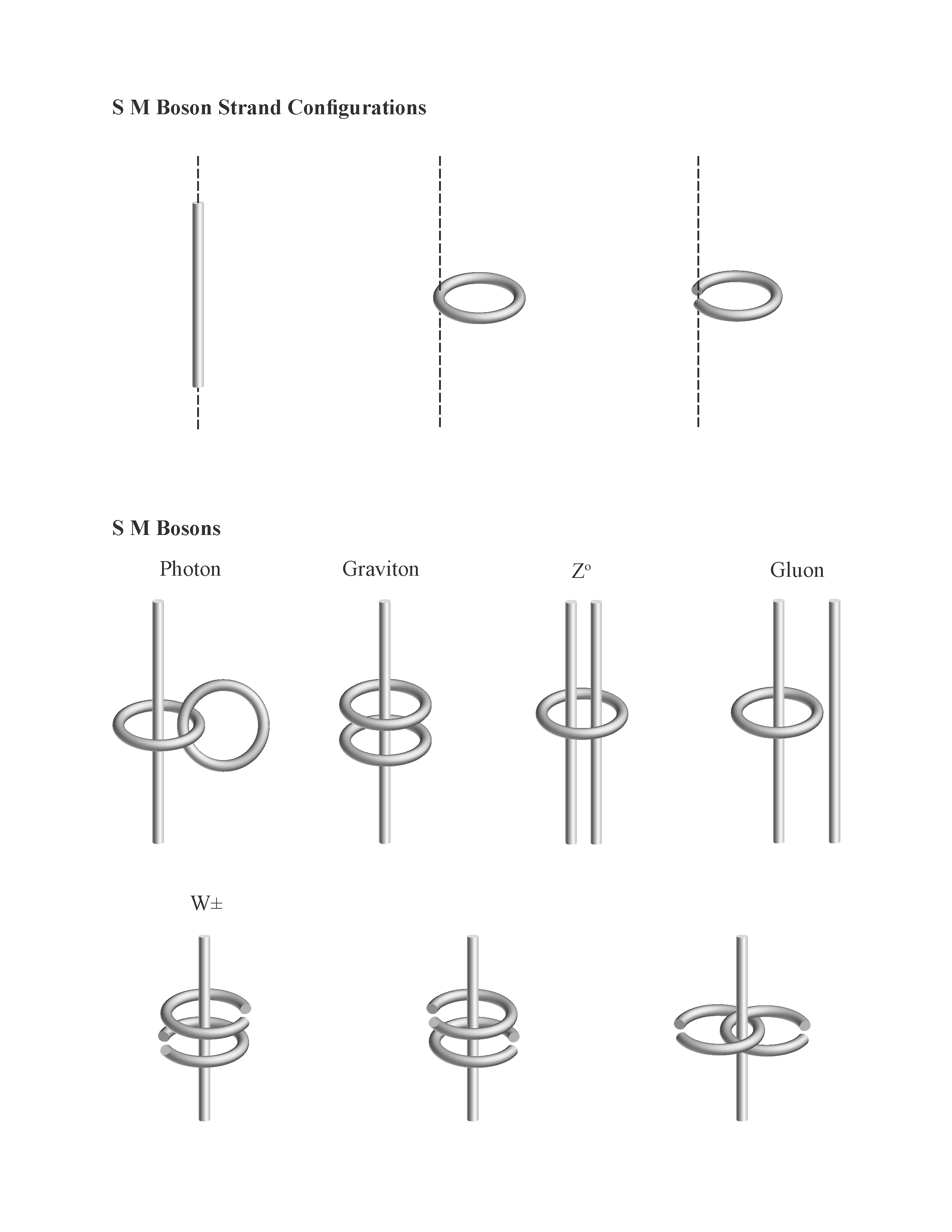}\caption{Preon strand configurations producing Standard Model bosons and the
resulting three preon boson states. Strands are shown as tubes for
ease of visualization.}
\end{figure}

\end{document}